\documentclass[%
 reprint,
 amsmath,amssymb,
 aps,
prl,
]{revtex4-2}

\usepackage{graphicx}
\usepackage{dcolumn}
\usepackage{bm}

\begin{document}


\title{Superconducting quantum computer: a hint for building architectures}

\author{Yutaka Tabuchi}
 \email{yutaka.tabuchi@riken.jp}
\author{Shuhei Tamate}%
\author{Shinichi Yorozu}%
\affiliation{%
 Center for Quantum Computing, RIKEN\\ 2-1 Hirosawa, Wako, Saitama, Japan 351-0198
}%

\date{\today}

\begin{abstract}
We discuss the scalability of superconducting quantum computers, especially in a wiring problem. The number of wiring inside a cryostat is almost proportional to the number of qubits in current wiring architectures. We introduce regularity, modularity, and hierarchy to an architecture design of superconducting quantum computers. The key to the wiring elimination is found in the quantum error correction codes having thresholds and spatial translational symmetry, {i.e.}, the surface code. We show a superconducting-digital-logic-based architecture and introduce a stacked heterogeneous structure of the quantum module.
\begin{description}
\item[Keywords] Quantum computing, Superconducting quantum circuits, Superconducting qubits.
\end{description}
\end{abstract}

\maketitle


\section{\label{sec:intro}Introduction}

The scalability of quantum computers (QCs) is an essential discussion in building architectures of QCs. Recently there have been proposed and demonstrated 50-qubit-sized superconducting QCs for near-term applications, {e.g.}, quantum chemistry and quantum simulations~\cite{bib:arute19,bib:kandala17}. Even with a 100-qubit-sized QC that solves a practical problem~\cite{bib:reiher17}, we keep working on increasing the number of qubits to thousands of thousands, given that we rarely obtain exact answers with a set of ``dirty'' qubits \cite{bib:footnote} having an error with a probability of c.a.~0.1$\sim$1.0~\%. For a reliable computation, we use a quantum-error-correction (QEC) code to make redundant ``clean'' qubits called logical qubits from sets of dirty qubits; a single logical qubit with a gate error probability of $10^{-8}$ requires a thousand qubits, for example. The tremendous required redundancy motivates us to expand the number of qubits and to integrate a system.  

Although intensive efforts have been made~\cite{bib:arute19,bib:rosenberg17} to increase the number of qubits, there is still room for a discussion about scalability. A practical obstacle in realistic QCs is a wiring problem where the number of wiring inside a refrigerator increases with the qubit numbers. Here, we discuss a hint to build a scalable architecture of superconducting QCs by figuring out the fundamental difference between conventional and quantum logic gates.

\section{\label{sec:gates}Conventional logic and quantum logic gates}

Conventional logic gates are restoring logic~\cite{bib:hillis15}. The essence of restoring logic gates is the suppression of voltage fluctuation propagation from the input to the output.  Each logic gate has thresholds, and a voltage with an error at input recovers to a well-defined value at the output pins using discrimination based on the threshold values. Thus, they are cascadable, and a deep combinational logic gate is made possible. 

On the other hand, quantum gates accept analog values at the input, which is the qubit wavefunction represented in a complex vector form.  We perform the quantum gates as an action to the analog input values; they are implemented using an analog control voltage waveform from room-temperature electronics. Because the analog values of coherent weights in the wavefunction are important for interference, we don't expect truncation of and unexpected change of the values due to noise for practical applications. In realistic cases, the quantum gates degrade due to noise in the control voltage waveform. The noise may originate from the crosstalk among control lines and waveform deformation to which circuit parameter variation contributes. As quantum gates have no error resilience that we have seen in the conventional logic, they cannot be restoring logic gates, {i.e.}, even $-$50~dB crosstalk in the control waveform become a bottleneck in a computation. Furthermore, quantum circuits are only combinational logic gates and lack sequential logic gates indispensable for the control circuits. The quantum circuits are supposed to work together with control circuitry, often located outside a refrigerator or at a higher temperature stage in the cryostat, demanding an enormous number of wiring between the chip and controller.

\section{\label{sec:threey}Regularity, Modularity, and Hierarchy in Quantum Computers}

We often rely on regularity, modularity, and hierarchy of components to build complicated systems like cars, aircraft, and computers~\cite{bib:harris12}. Regularity demands uniformity of elements to assemble a system simpler, {i.e.}, making an effort to reduce dedicated designs. Whereas conventional restoring logic gates are robust against variation of circuit parameters, quantum gates lack the regularity because they need individual control waveform calibration to all qubits to compensate for the shift in circuit parameters. The absence of regularity requests many wiring between the chip and controller in current QCs.  Modularity requires that the components have well-defined functions and interfaces so that we can readily assemble them. Quantum gates are analog elements. The connectivity fails as they have analog input and output without any error margin. Hierarchy is seen in a system that is dividable into modules which are consist of elementary sub-modules. In the hierarchical architecture, an upper module has fewer I/O pins than the sum of those of the lower modules, enabling us to control the system with simplified input and output, {e.g.}, an instruction set. Although arithmetic operations are possible in quantum circuits, it is hard to design state-machine-based control circuits due to the lack of sequential logic gates. The difficulty demands a gate-level control of all qubits using room-temperature electronics.

\section{\label{qec}Quantum Error Correction and Surface Code}

The use of QEC plays an essential role in not only alleviating errors due to noise and crosstalk but also realizing regularity, modularity, and hierarchy. QEC codes having error thresholds, {e.g.}, surface codes, provide identical logical qubits on the code as long as the error probability of dirty qubits is lower than the threshold value. In other words, the variation of the circuit parameters is allowable to the threshold value. Thus, they provide the regularity of quantum gate operations in the code space. The QEC codes also give an error margin to the action to the dirty qubits. The existence of the threshold induces robustness against the control and provides the margin; it drastically improves cascaded operations among modules.  Hierarchy of the control in the gate-level may avoid the difficulty in I/O wiring. However, even if the dirty qubits are identical, simple multiplexers (MUX) and demultiplexers (DEMUX) do not suit the QEC protocol in terms of control bandwidth. We simultaneously control all the qubits to detect errors in repetitive refresh operations and concurrently transmit the measurement outcomes to the room-temperature electronics. The MUXs and DEMUXs are the devices that provide multi-fanout features by reducing the communication bandwidth, and they are not the devices that hierarchize the functions in modules. 

A hint for the hierarchy is in the QEC itself. Surfaces codes~\cite{bib:fowler12} provide QEC features with thresholds on a two-dimensional grid of dirty qubits with nearest-neighbor interactions. As the code has spatial translation symmetry, the number of control patterns in the refresh operations is limited, leading to a drastic decrease in the control bandwidth. Logical gates to the clean qubits are inserted into the refresh operations every a few tens cycles for execution. The lattice surgery for the logical gates~\cite{bib:horsman12} requires a limited number of control operations; we are released from the individual control of qubits, leading to the suppression of the control bandwidth. The number of wiring is reduced using dedicated splitter circuits that switch the on-off combination depending on operation codes, {e.g.}, refresh, logical gates, and measurements on the code. We can regard the splitter circuit as an operation decoder (op-DEC) in micro-architecture, which installs a set of simple logical operations, and off-loads the logical gate implementation to the hardware. Besides, QEC decoder circuits at low temperatures~\cite{bib:holmes20,bib:ueno21} are ideal MUXs for the measurement outcomes. The high-throughput consecutive refresh outcomes are processed in the online decoder circuit to find the spatial and temporal error locations. Through the process, dirty qubits are encapsulated, and the only information of clean qubits is available from the outside of a cryostat. Hence, the surface code itself contributes to providing a hierarchy that reduces the I/O wiring.

\section{Implementation}
As quantum circuits lack sequential logic gates, we employ superconducting digital circuits available at low temperatures for the op-DEC and QEC decoder~\cite{bib:holmes20,bib:ueno21}. A tileable architecture is promising to reinforce the scalability of the system having regularity. The qubit and readout circuit pattern is repeated and tiled to eliminate the pattern-specific design. The splitter and surface code decoder circuitries are then placed on top or bottom of the qubit chip (Figs.~1(a), (b), and (c)) to scale the system in the two dimensions. The interface between the op-DEC and the qubit chips and those between the measurement and qubit substrates and between the measurement and QEC decoder chips need $2N$ and $N$ wirings (Fig.~1(b)), respectively, where $N$ is the number of qubits. The factor of 2 comes from the wiring to the control and readout circuits per qubit. It may also be better to control and measure the qubits from only one side of the qubit module, as in a conventional semiconductor logic fabrication process (Fig.~1(c)). The high-density wiring is accomplished using flip-chip bonding~\cite{bib:rosenberg17} shown in Fig.~1(d). The wiring number $M$, in contrast, significantly diminishes from the room temperature to the op-DEC. In superconducting QCs using fixed-frequency transmon qubits and cross-resonance gates for two-qubit gates, we can define a unit lattice of control having 20 dirty qubits ($k = 20$ in Fig.~1(b) and (c)). During the refresh operation, only 40 independent control and readout waveforms are necessary. Whereas the combination of logical gates is highly application-dependent, they may not have spatial translation symmetry. As a result of a trade-off for hardware off-loading, in that case, we need to either serialize the actions in time or design a complex splitter circuit having many op-codes.

\begin{figure*}
    \centering
    \includegraphics{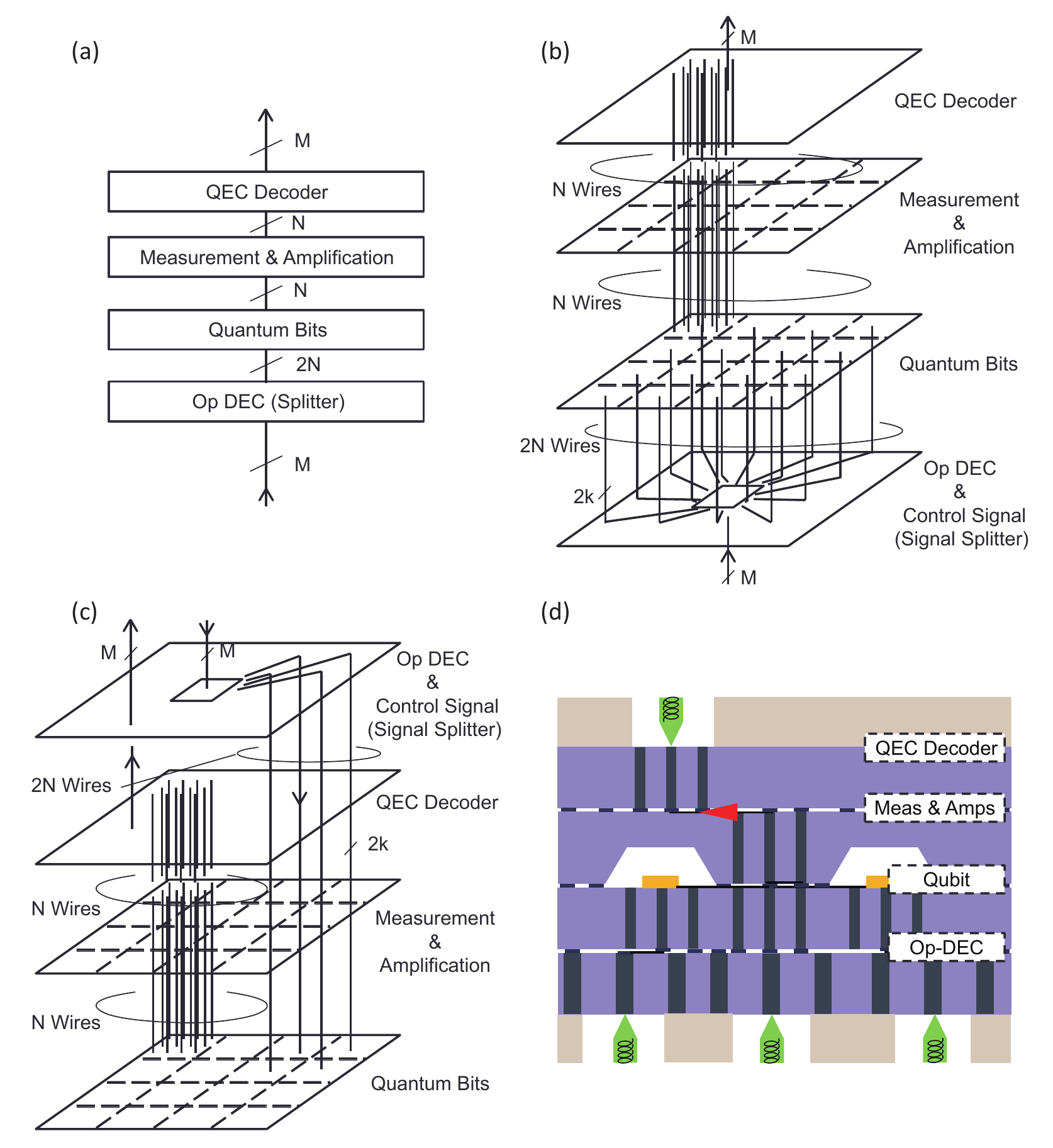}
    \caption{\label{fig:arch}(a) Four stacked substrates include control (op-DEC), qubit, readout, and signal processing (QEC decoder) functions located in a milli-Kelvin environment. The number of wiring from/to the stacked modules $M$ is far less than the number of qubits $N$. (b) (c) A unit lattice of control is depicted using dashed lines on qubit and measurement substrates. The control signal from the top/bottom part is expanded and distributed over the qubit chip. (d) High-density coaxial structures at the interfaces using through-silicon vias and micro-bump bondings. A coaxial waveguide with spring pin contacts is the I/O interface of the module communicating to the outside}
\end{figure*}

There are assumptions to realize QCs based on the proposed architecture. The dirty qubits need to be uniform enough for the threshold QEC operation. The superconducting digital logic can control qubits sufficient enough. The QEC decoder needs to be operable in low-power dissipation. Although there are many technical difficulties, building a concrete architecture with the hint above and developing relating technologies pave the path to practical quantum computation.

\section*{Acknowledgment} This work is supported by JST ERATO (No.~JPMJER1601) and by MEXT Q-LEAP (No.~JPMXS0118068682).


\begin{thebibliography}{99}
\bibitem{bib:arute19} F.~Arute \textit{et~al.}, ``Quantum supremacy using a programmable superconducting processor'', \textit{Nature}, vol.~574, pp.~505, October 2019.  
\bibitem{bib:kandala17} A.~Kandala \textit{et~al.}, ``Hardware-efficient variational quantum eigensolver for small molecules and quantum magnets'', \textit{Nature}, vol.~549, pp.~242, September 2017.
\bibitem{bib:reiher17} 
M.~Reiher, N.~Wiebe, K.~M.~Svore, D.~Wecker, and M.~Troyer, ``Elucidating reaction mechanisms on quantum computers'', \textit{Proc.~Natl.~Acad.~Sci.~USA}, vol.~114, pp.~7555, July 2017.

\bibitem{bib:footnote} ``Dirty'' qubits are referred to as ``physical'' qubits in the quantum information science community. 

\bibitem{bib:rosenberg17} D. Rosenberg \textit{et~al.}, ``Solid-state qubits integrated with superconducting through-silicon vias'', \textit{npj Quantum Inf.}, vol.~3, pp.~42, July 2017.
\bibitem{bib:hillis15} W.~D. Hillis, \textit{The Pattern On The Stone: The Simple Ideas That Make Computers Work}, Basic Books, 2015, p.~15-18. 
\bibitem{bib:harris12} D.~Harris and S.~Harris, \textit{Digital Design and Computer Architecture}, Morgan Kaufmann, 2012, pp.~6-7. 
\bibitem{bib:fowler12} A.~G.~Fowler, M.~Mariantoni, J.~M.~Martinis, and A.~N.~Cleland, ``Surface codes: Towards practical large-scale quantum computation'', \textit{Phys.~Rev.~A}, vol.~86, p.~032324, September 2012. 
\bibitem{bib:horsman12} C.~Horsman, A.~G.~Fowler, S.~Devitt and R.~V.~Meter, ``Surface code quantum computing by lattice surgery'', \textit{New J. Phys.}, vol.~14, p.~123011, December 2012. 
\bibitem{bib:holmes20} A.~Holmes, M.~R.~Jokar, G.~Pasandi, Y.~Ding, M.~Pedram, and F.~T.~Chong, ``NISQ+: boosting quantum computing power by approximating quantum error correction'', ISCA '20: Proceedings of the ACM/IEEE 47th Annual International Symposium on Computer Architecture, 2020, pp.~556. 
\bibitem{bib:ueno21} Y.~Ueno, M.~Kondo, M.~Tanaka, Y.~Suzuki and Y.~Tabuchi, ``QECOOL: On-line quantum error correction with a superconducting decoder for surface code'', arXiv:~2103.14209.
\end{thebibliography}
\end{document}